\documentclass[aps,prl,reprint,showpacs,twocolumn,floatfix]{revtex4-1}
\usepackage{amsmath}
\usepackage{amssymb}
\usepackage{amsfonts}
\usepackage{makeidx}
\usepackage{graphicx}
\usepackage{bm}
\usepackage{epstopdf}

\newcommand{\bea}{\begin{eqnarray}}
\newcommand{\eea}{\end{eqnarray}}
\newcommand{\be}{\begin{equation}}
\newcommand{\ee}{\end{equation}}

\begin{document} 

\title{A pseudo-likelihood decimation algorithm improving the inference of the interaction network in a general class of Ising models}

\author{Aur\'elien Decelle$^{1}$ and Federico Ricci-Tersenghi $^{1,2}$}

\affiliation{
$^1$Dipartimento di Fisica, Universit\`a La Sapienza, Piazzale Aldo Moro 5, I-00185 Roma, Italy. \\
$^2$INFN-Sezione di Roma 1, and CNR-IPCF, UOS di Roma.
}

\begin{abstract}
In this Letter we propose a new method to infer the topology of the interaction network in pairwise models with Ising variables.
By using the pseudo-likelihood method (PLM) at high temperature, it is generally possible to distinguish between zero and non-zero couplings, because a clear gap separate the two groups.
However at lower temperatures the PLM is much less effective and the result depends on subjective choices, as the value of the $\ell_1$-regularizer and that of the threshold to separate non-zero couplings from null ones.
We introduce a decimation procedure based on PLM, that recursively sets to zero the less significant couplings, until the variation of the pseudo-likelihood signals that relevant couplings are being removed.
The new method is fully automated and does not require any subjective choice by the user.
Numerical tests have been performed on a wide class of Ising models, having different topologies (from random graphs to finite dimensional lattices) and different couplings (both diluted ferromagnets in a field and spin glasses). These numerical results show that the new algorithm performs better than standard PLM.
\end{abstract}

\maketitle

Recent years have seen a growing interest of the statistical physics community in the inverse Ising problem \cite{cocco2011adaptive, inf_aurell-11, ricci2011mean, nguyen2012bethe, cocco2012adaptive, ekeberg2013improved, raymond2012mean}. Although the problem is known since a long time ago under the name of Boltzmann machine learning \cite{ackley1985learning}, the renewed interest is linked to the large number of datasets coming from many different fields --- e.g.\ biology, physics, neuroscience --- that require a quantitative description. Building reliable models for these datasets has become therefore a fundamental problem \cite{BialekNature, WeigtPNAS, CoccoPNAS, mora2010maximum}. Given a dataset and a model, the inverse problem aims to find the parameters of the model that best fit the data. Among all possible models, the Ising one, although very simple --- it involves only pairwise interactions between discrete variables --- can take into account a wide range of phenomena. It is thus natural to develop inference methods for this statistical model, as initially done by Boltzmann machine learning \cite{ackley1985learning}.

A common approach in Bayesian inference is to infer the model couplings by maximizing the likelihood function. Unfortunately, the likelihood depends on the partition function which is extremely difficult to compute in general. Therefore we use the pseudo-likelihood approximation \cite{besag1975statistical,wainwright2010AnnalHigh} to perform inference on Ising models of reasonable size. This method has at least three advantages: the pseudo-likelihood function (PLF) can be maximized in polynomial time; the method is known to be exact in the case of infinite sampling \cite{inf_aurell-11}; as we will see in this Letter, the PLF can be used as an indicator of how good the reconstruction is, allowing one to decrease accurately the number of model parameters. 

We begin by introducing the pseudo-likelihood method (PLM) that has been already tested in inferring finite dimensional models and found to be very effective \cite{inf_aurell-11}. We also recall the $\ell_1$-regularization extension and the thresholding procedure for discriminating non-zero couplings. We discuss the problem of very large coupling appearing in the PLM and how we solve this issue without using the $\ell_1$-regularization. Finally, we describe the new decimation procedure to select accurately the model parameters and we show that the PLF can be used as a reliable likelihood. We conclude by showing that our new symmetrized PLM with decimation provides a very good inference of model couplings, better than standard PLM, in a wide class of Ising models.

Let us consider an Ising model with probability measure
$P(\underline{s})= Z^{-1}\exp\Big(\beta \sum_{i<j} J_{ij} s_i s_j + \beta \sum_i h_i s_i\Big)$,
whose parameters $\{J_{ij},h_i\}$ we want to infer, given a set of $M$ configurations $\{\underline{s}^{(k)}\}_{k=1\ldots,M}$ independently sampled from $P(\underline{s})$.
Bayesian inference prescribes to maximize with respect to the model parameters the likelihood of the data
$\mathcal{L} = \beta \sum_{i<j} J_{ij} \langle s_i s_j \rangle_D + \beta \sum_i h_i \langle s_i \rangle_D - \log(Z)$, where $\langle\cdot\rangle_D$ is the average over the data.
The maximum of $\mathcal{L}$ corresponds to parameters $\{J_{ij}^*,h_i^*\}$ matching the magnetizations and the correlations of the inferred model with those from the data.
There are two general methods to find the parameters $\{J_{ij}^*,h_i^*\}$ that fulfill these conditions. The first one is by finding the maximum of $\mathcal{L}$ using a gradient descent approach, but this requires to compute the partition function many times, which is impossible in practice. The second one, named Boltzmann machine learning \cite{ackley1985learning}, computes magnetizations and correlations by Monte Carlo methods and updates the parameters $\{J_{ij},h_i\}$ accordingly. Again this requires to run several long Monte Carlo simulations evaluating the average values with a precision good enough.

An important aspect in the inverse problem is therefore to find an approximation that provides a reasonably good estimate for the model parameters in a quick time (usually polynomial in the system size). A review of known approximations can be found in \cite{roudi2009statistical}, to which we should add the more recent methods, as those based on the Bethe approximation \cite{ricci2011mean,nguyen2012bethe}, the adaptive cluster expansion \cite{cocco2011adaptive} and the probabilistic flow method \cite{sohl2011new}. In this Letter we compare our new decimation technique based on the PLM against the $\ell_1$-regularization which is very much used to find the interaction topology \cite{wainwright2010AnnalHigh,inf_aurell-11,BentoMontanari}. Standard mean-field method will not be considered as in general they perform weakly on finite dimensional systems, neither the adaptive cluster expansion nor the probabilistic flow method, since they strongly depends on several subjective choices (thresholds, dynamics, etc...).

Instead of using the likelihood function, which is very hard to compute, we use the PLF, $ \mathcal{PL} = \sum_r \mathcal{L}_r$, where the index $r$ runs over all variables and the ``local'' likelihood functions are
$\mathcal{L}_r = M^{-1} \sum_{k=1}^{M} \log p(s_r^{(k)}| \underline{s}_{\setminus r}^{(k)})$,
with $p(s_r|\underline{s}_{\backslash r}) \equiv \Big[1+\exp(-2 \beta s_r (h_r + \sum_{j \neq r} J_{rj} s_j))\Big]^{-1}$ being the conditional probability of variable $s_r$ given the rest of the system.

The standard implementation of the PLM consists in maximizing each of the $N$ local likelihood functions $\mathcal{L}_r$ separetely, thus getting two different estimates for each coupling $J_{ij}$: a first one $J_{ij}^{*i}$ from the maximum of $\mathcal{L}_i$ and a second one $J_{ij}^{*j}$ from the maximum of $\mathcal{L}_j$ (hereafter we assume $h_i=0$ and $J_{ij} \in \{0,1\}$ for the ease of presentation). Since the Ising model has symmetric couplings, the final estimate for the coupling $J_{ij}$ is then obtained by taking the average, $J_{ij}^* = (J_{ij}^{*i}+J_{ij}^{*j})/2$.

This method has been confronted against mean-field methods for the SK model in \cite{inf_aurell-11} and it clearly estimates the couplings much better at low temperatures.
For sparse models, its ability to infer correctly the interaction network (i.e.\ which couplings are non-zero) is largely improved by the use of the $\ell_1$-regularization \cite{wainwright2010AnnalHigh}, thus maximizing the local functions $\mathcal{L}_r^{\ell_1} = \mathcal{L}_r + \lambda \sum_{j \neq r}|J_{rj}|$, with a suitably chosen (and not too large) $\lambda$ regularizer. A further improvement in inferring the model topology has been achieved \cite{inf_aurell-11} by setting to zero all couplings whose estimate is below a threshold, $|J_{ij}^*|<\delta$ (but the choice of $\delta$ is delicate, as we discuss below).

Using the standard PLM we observe that, in difficult situations (e.g.\ when $M$ is not large enough and temperature is low) some couplings are largely overestimated, $|J_{ij}^*| \gg 1$. In those cases, the inferred couplings are not symmetric ($J_{ij}^{*i} \neq J_{ij}^{*j}$) and only one of the two estimates is very large. The origin of this problem is to be found in the data information content, that is sometimes very poor around some variable $s_i$: in that case the estimates $J_{ij}^{*i}$ are strongly unreliable (think e.g.\ to what happen if in the $M$ samples $s_i$ and its neighbors were almost always perfectly aligned). This problem is often partially solved by using the $\ell_1$-regularization: the  regularization parameter $\lambda$ adds a penalty on non-zero couplings and therefore prevents a coupling from being too large. A drawback of this approach is the tendency to underestimate globally the couplings and is therefore not completely satisfying. In this Letter, we choose a different solution: by maximizing the PLF $\mathcal{PL}$ (rather than the $N$ functions $\mathcal{L}_r$ separately) we look for a compromise, where the estimate $J_{ij}^*$ must be such that both $\mathcal{L}_i$ and $\mathcal{L}_j$ are reasonably large. The advantage of this maximization is that it provides a {\em unique} estimate for each coupling, which is in general of the right order of magnitude (unless the information content of the data is poor in a wide region). The PLF can be maximized by a standard Newton method, while for the $\ell_1$-regularized functions we use the one of Ref.~\cite{koh2007interior}.

In inference problems, the ``model selection'' is the ability to reduce the number of model parameters in order to keep only the essential ones (by Occam's razor rule the simplest model fitting a dataset is the one to be preferred). This model selection is both crucial to identify the structure underlying a dataset (e.g. the interaction network or the model topology) and to improve the quality of the coupling estimates (relevant couplings are better inferred after excluding the insignificant couplings from the model).

\begin{figure}
\includegraphics[width=\columnwidth]{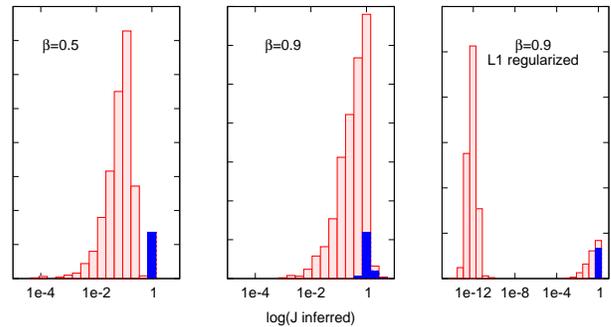}
\caption{(color online) Histograms of couplings inferred by PLM in a typical sample of 2D ferromagnetic Ising model with 30\% dilution ($M=4500$). Left: at $\beta=0.5$ a clear gap identifies couplings in $\bm{J}_1$ (drawn in dark blue). Center: at $\beta=0.9$ there is no gap and couplings in $\bm{J}_1$ can not be identified. Right: with $\ell_1$ regularization ($\lambda=0.01$) a gap is present, but does not allow to recover perfectly the topology.}
\label{fig_COUP}
\end{figure}

A common approach in the inverse Ising problem is to choose which are the null couplings by putting a cut-off on their estimated values. This is a reasonable choice as long as a clear gap separates the estimates of null couplings ($\bm{J}_0$) from those of non-zero couplings ($\bm{J}_1$), as it happens in the left panel of Fig.~\ref{fig_COUP}. However at a lower temperature the gap is completely absent and a good $\delta$-thresholding to split $\bm{J}_0$ and $\bm{J}_1$ is impossible (central panel of Fig.~\ref{fig_COUP}). For sparse models, the use of the $\ell_1$-regularization may induce a new gap (see right panel of Fig.~\ref{fig_COUP}), but this $\ell_1$-induced gap does not always separates correctly $\bm{J}_1$ from $\bm{J}_0$ couplings and the topology can not be correctly recovered. Moreover the estimates of $\bm{J}_1$ are systematically smaller than the true values, and this is unavoidable when using the $\ell_1$-regularization. Finally, there is no consensus on how to choose the $\lambda$ value.

We propose a new method for inferring non-zero couplings $\bm{J}_1$ that solves all the above problems. Our idea is to recursively set to zero couplings which are estimated very small by the PLM: we always maximize $\mathcal{PL}$ so as to avoid too large couplings and the bias due to the $\ell_1$-regularization. At each decimation step we set to zero a finite fraction $\rho$ of the remaining couplings, as such the total number of steps is $O(\log N)$ and the PLM+decimation algorithm is competitively fast. This fraction $\rho$ is actually the only choice left to the user and the results are largely independent on it (in our tests we have used $\rho \le 0.05$).
Setting couplings to zero gradually is equivalent to using an adaptive threshold with a very small $\delta$ value: so our new method should perform better than any standard thresholding procedure, especially because it does not require the existence of a gap in the inferred couplings (thus avoiding the use of $\ell_1$-regularization that produces biased estimates and a strong $\lambda$-dependence).

The stopping criterion for the decimation procedure is based on the behaviour of the PLF. Indeed, we expect that, as long as the decimation procedure sets to zero couplings in $\bm{J}_0$ which are unnecessary to fit the data, the PLF should not change significantly. On the contrary, the pruning of a coupling in $\bm{J}_1$ should produce a drastic decrease in the PLF value. This expected behavior is confirmed by the numerical simulations. In practice we would like to stop the decimation at the point where the PLF variation, $\Delta \mathcal{PL}/\Delta n$ with $\Delta n$ being the number decimated couplings in the last step, goes from `small' to `large' values. To make these two adjective quantitative we can compute the overall mean PLF variation during the decimation, that is the change in PLF between the fully connected model [where $\mathcal{PL}$ is maximized over all the $N(N-1)/2$ possible couplings and takes value $\mathcal{PL}_\text{max}$] and the model of independent variables [no couplings left by the decimation and PLF equal to $-N\log(2)$]. The mean PLF variation is thus equal to $(\mathcal{PL}_\text{max}+N\log(2))2/(N(N-1))$ and we propose to stop the decimation where the PLF variation reaches this mean value, that should separate `small' from `large' PLF variations. In practice it is more convenient to define the stopping point as the maximum of the {\em tilted PLF} (tPLF)
$\mathcal{PL}^\text{tilted} \equiv \mathcal{PL} - x \mathcal{PL}_\text{max} + (1-x) N \log(2)$,
where $x$ is the fraction of non-decimated coupling. It is easy to check that $\mathcal{PL}^\text{tilted}=0$ both before starting the decimation ($x=1$) and on a model with no coupling ($x=0$). In the interval $[0,1]$ a maximum appears if correlations are present in the dataset.

\begin{figure}
\includegraphics[width=0.8\columnwidth]{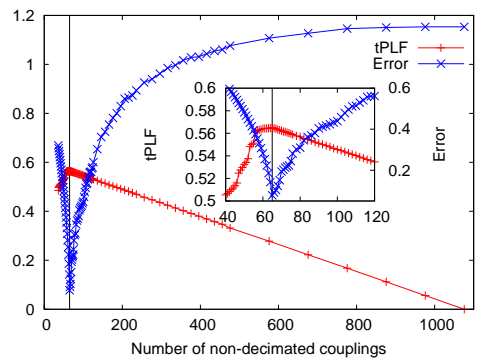}
\caption{Same model as Fig.~\ref{fig_COUP} with $\beta=0.9$ and $M = 4500$. The tPLF increases as the number of non-decimated couplings is reduced until it exhibits a maximum. In this case the maximum corresponds to a complete recovery of the graph topology and therefore to a small reconstruction error. The inset is a zoom on the maximum region.}
\label{fig_DECI}
\end{figure}

In Fig.~\ref{fig_DECI} we show, for a case where inference is difficult, the tPLF as a function of the number of non-decimated couplings and the corresponding error in inferring couplings, defined as
$\epsilon=\sqrt{\sum_{i<j} (J_{ij}-J_{ij}^*)^2/\sum_{i<j} J_{ij}^2}$.
The zoom in the inset clearly shows that the maximum in the tPLF does corresponds to the minimum in the inference error. Please notice also as the density of data points along the curves changes, because we have decreased the value of $\rho$ during the decimation in order to spend less time in the initial part (which is easy) and have denser points close to maximum (and thus improve its location).

\begin{figure}
\includegraphics[width=0.8\columnwidth]{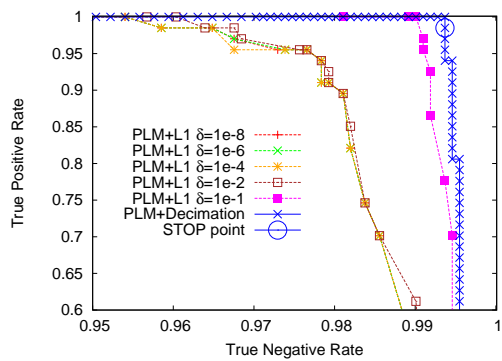}
\caption{(color online) ROC curves for a typical sample of the 2D ferromagnetic Ising model with $30\%$ of dilution at $\beta=1.0$ and $M=4500$ (difficult case). The upper right corner corresponds to the perfect reconstruction. For PLM+$\ell_1$ curves are drawn by varying $\lambda$ and they almost coincide if the threshold $\delta$ is correctly chosen in the gap, $\delta \in [10^{-8},10^{-2}]$. The better value $\delta=0.1$ can not be chosen without previous knowledge about the topology. The PLM+decimation is clearly inferring topology better, even if PLM+$\ell_1$ is finely tuned over $\lambda$ and $\delta$.}
\label{fig_ROC}
\end{figure}

Our new inference algorithm (PLM+decimation) can be very well compared to the standard PLM with $\ell_1$-regularization and $\delta$-thresholding by plotting the corresponding ROC curves (see Fig.~\ref{fig_ROC}). Each ROC curve is obtained by plotting parametrically the true positive rate (i.e.\ couplings in $\bm{J}_1$ inferred as non-zero divided by the total number of couplings in $\bm{J}_1$) versus the true negative rate (i.e.\ couplings in $\bm{J}_0$ inferred as null divided by the total number of couplings in $\bm{J}_0$). The ROC curve for an exact inference method run on noiseless data would pass through the upper right corner. In general an inference method is better the larger the area below the ROC curve. In the present case, the ROC curves for the standard PLM have been drawn by varying $\lambda$ at fixed $\delta$, while for the PLM+decimation the ROC curve has been drawn by varying the fraction of decimated couplings. Clearly the new method is outperforming standard PLM, even if the latter were optimized over $\lambda$ and $\delta$. To this respect, it is worth noticing that $\delta$ values in the range $[10^{-8},10^{-2}]$ do actually fall in the gap (see lower panel of Fig.~\ref{fig_COUP}) and lead to very similar ROC curves, while $\delta=0.1$ is outside the gap and would be impossible to choose that value without knowing in advance the topology we are looking for. So the better ROC curve with $\delta=0.1$ is practically unachievable and in general we believe that optimizing over $\delta$ should not produce any sensible improvement as long as $\delta$ is correctly chosen in the gap (to this respect the improvement obtained in Ref.~\cite{inf_aurell-11} makes us suspect that a too large $\delta$ value was chosen based on the previous knowledge of the topology). So a fair comparison of our new method with the standard PLM should be made by choosing $\delta \le 10^{-2}$ and the improvement is then very large. The stopping point selected by maximizing the tPLF is shown by a large dot in Fig.~\ref{fig_ROC}, and is indeed the closest to the upper right corner (full topology recovery).

\begin{figure}[t]
\includegraphics[width=0.89\columnwidth]{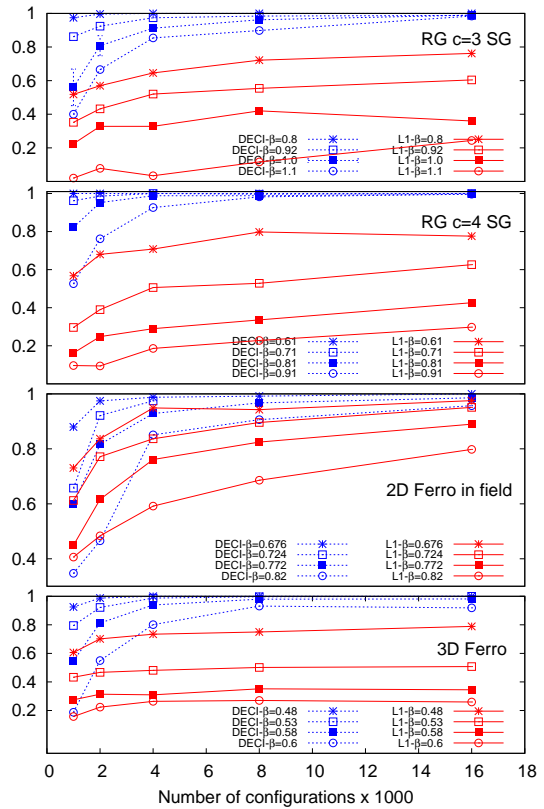}
\caption{(color online) Neighbourhood reconstruction for four different models as a function of the number of samples. Dashed blue lines are results by the new decimation algorithm, while full red lines are by PLM+$\ell_1$. Each point is the average over $10$ samples. Top to bottom: spin glass models, $J_{ij}=\pm1$, on random graph, $N=50$, with average degree $c=3$ and $c=4$; diluted ferromagnetic models, in 2D, $N=49$, with 30\% of dilution and a random Gaussian field with zero mean and variance $0.3$, and in 3D, $N=64$, with 55\% of dilution. The decimation algorithm is always better and performs particularly well on random graphs.}
\label{fig_GSUCC}
\end{figure}

We applied the standard PLM+$\ell_1$ and our new decimation algorithm to other models to compare their ability in recovering the topology of the underlying graph. We studied two diluted ferromagnetic models (in 3D with 55\% dilution and in 2D with 30\% dilution and random fields) and two spin glass models ($J_{ij}= \pm 1$) on Erd\"os-Renyi random graphs with average degrees $c=3$ and $c=4$. To measure the performance of the inference methods we shown in Fig.~\ref{fig_GSUCC} the fraction of successfully reconstructed neighbourhoods
as a function of the number of samples $M$.
It is clear that the new decimation algorithm outperforms standard PLM+$\ell_1$ in reconstructing any model topology.
Please consider that for PLM+$\ell_1$ method $\lambda$ is chosen as to minimize the distance on the ROC curve from perfect reconstruction (the point closer to the upper-right corner) and therefore this choice is made knowing the topology. The threshold $\delta$ was always set in the gap (typically $\delta \sim 10^{-6}$) and we observe that this method does not manage to separate $\bm{J}_0$ from $\bm{J}_1$, even at high sampling.
In order to keep the figure more readable, the uncertainties on data points are drawn only for one dataset in the upper left panel (the others errors being similar or smaller). These uncertainties have been computed by resampling the configurations passed as input to the inference algorithm.

Compared to the methods presented in \cite{LeCun, Hassibi}, our algorithm performs similarly and in a quicker time, since it does not need to compute the inverse of the Hessian.

In conclusion, we have presented a new method for inferring the interaction topology of Ising models which is based on the PLM and a decimation procedure that recursively sets to zero couplings which are inferred as the weakest.
The procedure is fully automated (apart from the choice of $\rho$ which is mostly irrelevant for the results) and provides a unique answer to the inverse Ising problem and the corresponding model selection problem.
Execution times are comparable to those of standard PLM (i.e.\ polynomial in system size), apart from an extra $O(\log N)$ multiplicative factor (but remind that maximization without the $\ell_1$-regularizer is easier).
As the standard PLM, also our new method is exact in the limit of very large number of samples $M$. For finite (and small) values of $M$ we have run extensive numerical tests on a wide class of Ising models, with different topologies (from random graphs to finite dimensional lattices) and different couplings (both diluted ferromagnets in a field and spin glasses).
The results show that the new algorithm performs much better than standard PLM with $\ell_1$-regularization and $\delta$-thresholding, which was considered among the best inference techniques available.

\begin{acknowledgments}
This research has received financial support from the Italian Research Minister through the FIRB project No. RBFR086NN1.
\end{acknowledgments}

\end{document}